# An accurate IoT Intrusion Detection Framework using Apache Spark

Mohamed Abushwereb
*Department of Computer Science*
*Princess Sumaya University for Technology*
Amman, Jordan
mohamed.abushwereb@gmail.com

Mohammad Almseidin
*Department of Information Technology*
*University of Miskolc*
Miskolc, Hungary
alsaudi@iit.uni-miskolc.hu

Mouhammd Al-kasassbeh
*Department of Computer Science*
*Princess Sumaya University for Technology*
Amman, Jordan
m.alkasassbeh@psut.edu.jo

Muhannad Mustafa
*Department of Computer Science*
*Princess Sumaya University for Technology*
Amman, Jordan
muhannad.a.mustafa@gmail.com

**Abstract** — The internet has caused tremendous changes since its appearance in the 1980s, and now, the Internet of Things (IoT) seems to be doing the same. The potential of IoT has made it the center of attention for many people, but, where some see an opportunity to contribute, others may see IoT networks as a target to be exploited. The high number of IoT devices makes them the perfect setup for staging denial-of-service attacks (DoS) that can have devastating consequences. This renders the need for cybersecurity measures such as intrusion detection systems (IDSs) evident. The aim of this paper is to build an IDS using the big data platform, Apache Spark. Apache Spark was used along with its ML library (MLlib) and the BoT-IoT dataset. The IDS was then tested and evaluated based on F-Measure (f1), as was the standard when evaluating imbalanced data. Two rounds of tests were performed, a partial dataset for minimizing bias, and the full BoT-IoT dataset for exploring big data and ML capabilities in a security setting. For the partial dataset, the Random Forest algorithm had the highest performance for binary classification at an average f1 measure of 99.7%, as well as 99.6% for main category classification, and an 88.5% f1 measure for sub category classification. As for the complete dataset, the Decision Tree algorithm scored the highest f1 measures for all conducted tests; 97.9% for binary classification, 79% for main category classification, and 77% for sub category classification.

## I. INTRODUCTION

Since the internet was established in the 80s and 90s, economies and businesses have changed at tremendous speed, given the new opportunities this technology has offered. Now we are at the edge of a new transformation; where before the internet was only a cluster of computers and phones, now ever-smaller controllers have enabled many devices, sensors, and 'things' in general to connect over the internet as well, ushering in the age of the IoT [1]. As the internet has opened new opportunities in the past, the IoT now also shows great promise in many areas, such as the industrial, health, and military sectors, to name a few [2]. As such, the IoT has grown into a hundred billion-dollar industry with many applications, as shown in Fig 1, some already serving their use such as industrial applications with a 26.4% market share, healthcare applications with a 22% market share, and smart homes with a 15.4% market share. Other applications are still being tested and explored, such as smart cities with a market share of 28.6%, and connected cars with 7.7%. With this enthusiasm towards IoT and the capabilities it offers [3], it is estimated

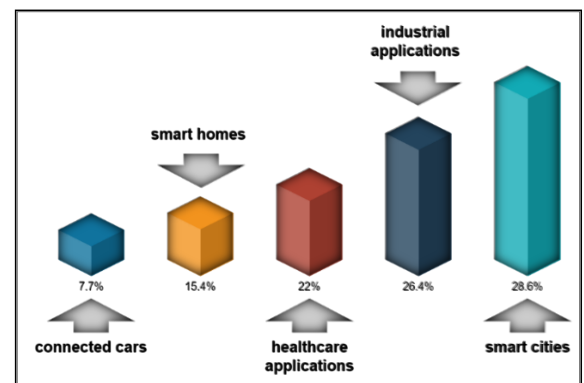

Fig 1 IoT Market share in 2019 by sector

that there will be more than 25 billion devices connected by the end of 2020 [4].

While this is a fascinating prospect for business owners, engineers, and consumers alike, the massive investments into these devices' variety and number attract another kind of people with more sinister intentions—hackers [5]. A hacker with access to an IoT network can use a large number of devices to flood another system with requests, rendering its services inaccessible altogether, an attack commonly known as a DoS. These attacks can use an IoT network owned by the hacker, or a network obtained by unauthorized access. The devices in these networks are then used as zombies or as a botnet to attack servers and other networks. One of these recent DoS attacks is the infamous Mirai attack of 2016, which had generated a volume of over 600 Gbps [6]. The dangerous implications of this attack and others have highlighted the need for enhanced cybersecurity measures concerning IoT devices [7] [8] [9].

One of the essential cybersecurity applications with regard to securing networks is the IDS. While it has been in use since the late 80s, the ever-increasing methods of malicious attacks have caused IDSs to evolve as well [10] [11]. Early IDSs used a signature-based method, where all known attack patterns (or signatures) were stored in a database and compared to incoming traffic to detect attacks [12]. After the number of DoS attacks multiplied, it became unfeasible to store all the signatures in a database, and the anomaly-based method was introduced, using AI to predict new attacks. Prediction is achieved using ML algorithms that take in data of previous or known attacks as training data come up with accurate predictions of future attacks, and finally measure these predictions against new traffic [13].

IDSs that use ML have been used in traditional networks and have been shown to be very efficient and accurate [14] [15]. However, there still aren't enough studies centered around ML in IoT networks for two reasons: the first is that ML algorithms need reliable datasets to train their models, and most published datasets are only from traditional networks. The second reason is that IoT networks usually work in real-time ecosystems with instant response and large quantities of data. Researchers have tried to use traditional ML techniques to get around these constraints while maintaining high accuracy, but the big data generated by IoT networks always slowed the process. A framework is needed that can process big data, and can be used in conjunction with ML techniques and libraries to classify incoming traffic into an attack or normal data. Apache Spark is one such notable framework [16].

The aim of this paper is to develop a big data model in Apache Spark that can detect IoT network attacks for binary classification and multiclass classification. In addition to big data, ML classifiers are also used in classifying network traffic. The BoT-IoT dataset, developed by Cyber Range Lab of UNSW Canberra Cyber, will be used for training and testing the IDS. As for evaluation, the system will be rated based on the detection and classification of different attacks.

This paper is structured as follows: first, a firm groundwork is laid out in section 2. Section 3 then goes over the methodology of this paper in detail. Results and discussion of the experiments are then spread out over section 4.

## II. RELATED WORK

The literature review was divided into three main categories based on the combination of these elements. The first section is for papers that combine AI and IoT, the second section is for AI and big data, and finally, the third section contains only research that has a combination of all three.

### A. AI-based detection techniques for IoT networks

In the following section, an overview of previous research regarding the use of AI in securing IoT networks is given; this includes different disciplines of AI, such as ML and Deep Learning (DL).

One of the sectors IoT is being adapted for is the military sector. The author in [17] attempts to detect malware stowed away in the data of IoT networks in military defense systems. The author achieves this through the use of deep eigenspace learning and operational code sequences, where data is transformed into a vector space representation, and deep eigenspace learning is used to recognize malicious software. The solution was tested against junk code insertion attacks and evaluated based on four criteria: accuracy, precision, recall, and f-measure. The results were also compared to two previous works with the conclusion that the deep eigenspace learning approach had better results, with an accuracy rate of 99.68%, a precision of 98.59%, recall of 98.37%, and f-measure of 98.48%. However, the dataset used in the paper was of the author's design and had a limited number of malware, bringing into question the quality and validity of the dataset.

Because IoT devices are easier to compromise, the number of DoS attacks on IoT networks has been increasing. To minimize these attacks, [18] introduced a new way to detect anomalies in IoT networks using a deep auto encoder, which was taken from an already compromised network that was attacked by Mirai and BASHLITE. The evaluation of this approach was comprised of True Positive Rate (TPR), FPR, and attack detection time, which resulted in a 100% TPR, an FPR of $0.007 \pm 0.01$, and a detection time of $174 \pm 212$ milliseconds. But these results only account for two types of DoS attacks.

DL is also used in cybersecurity applications for IoT, as demonstrated in [19]. The authors used a Bi-directional Long Short-Term Memory (LSTM) Recurrent Neural Network (RNN) to detect attacks and was evaluated using six criteria: accuracy, precision, recall, f-measure, miscalculation rate, and detection time. The neural network gained an accuracy of 95.7% but wasn't compared to other solutions of the same type to ensure its effectiveness.

Some research focuses on securing IoT networks at the application layer, such as [20], which proposes a framework for IoT networks based on Software-Defined Networks. The evaluation consisted of eight criteria: TP, FP, TN, FN, precision, recall, False Discovery Rate, and FNR. Although the framework had a precision rate of more than 94%, the dataset used (the KDD99 dataset) is old, containing attacks from 1999. The dataset also has too small a sample of attacks, such as DoS and reconnaissance attacks. If a newer dataset was used, the results of this paper would have been more reliable.

The work represented in [21] took a different approach to previous studies. The author addresses the problems surrounding an IoT network's confidentiality, integrity, privacy, and availability. This is done by a model that detects attacks within the IoT network itself. The following metrics were considered to evaluate the model: accuracy, precision, recall, and detection time. Results showed promise in speed and accuracy when the NSL-KDD dataset was used and accuracy rated more than 95%. However, the model's performance deteriorated when switched to a newer dataset, namely the UNSW-NB15 dataset, in which detection time increased significantly, and accuracy dropped below 95%, showing that this model only works as intended with older datasets such as NSL-KDD, an extension of the KDD99 dataset that is much older than UNSW-NB15.

Another DL approach was taken to build a detection method for IoT networks [3], namely the Feed-forward Neural Network (FNN) DL method. FNN was used with the BoT-IoT dataset, which was released in 2018 and contained five categories of attacks, to detect attacks in both binary classification and multiclass classification scenarios. The approach achieved high accuracy in binary classification with a rate of 99.9% and was able to separate normal and malicious data patterns in multiclass classification with a rate of 98%. Still, it failed at classifying the attacks themselves and even had an accuracy rate of no more than 88.9% in one of the attack types. Furthermore, binary classification was not tested on all attacks simultaneously, but rather one type of attack at a time, putting into question this approach's ability to successfully detect attacks of multiple types in a binary classification scenario.

Where most attempts choose between signature-based and anomaly-based detection, study [22] sought to make use of both in a Hybrid IDS. Detection happened across two phases. The first was signature-based and used the C5 algorithm to speed up the process and decrease the second phase's load. The second phase was anomaly-based and used a one-class SVM classifier to thoroughly go through the data marked as dubious. This Hybrid IDS managed an accuracy rate of 99.97%, but although the initial idea was promising, its application wasn't accurate, as the subset of data that was taken as a test sample from the dataset was highly imbalanced. Moreover, no preprocessing processes were implemented, which led to a high rate of false-positive predictions.

Another paper that used the BoT-IoT dataset is research [23], which used the dataset to train two models with the aim of measuring the effect of adversarial attacks on IoT networks. The first model was built as an FNN, while the second was built as a Self-normalized Neural Network (SNN). Finally, the BoT-IoT dataset was then used to test both models with and without adversarial attacks. The study concluded the SNN model was more efficient even after adversarial attacks were added to the dataset. However, it did not show the full results of the effects that adding adversarial attacks had on both models and their ability to classify other kinds of attacks.

Authors [24] compared the accuracy and efficiency of seven different algorithms in detecting DoS attacks on IoT networks. Two of these algorithms were hybrids that used MultiSchemes and Voting, the first of which integrated the Averaged one-Dependence (A1DE), and the second which integrated the Averaged two-Dependence (A2DE). When all algorithms were tested using the BoT-IoT dataset, the hybrid model came first in terms of accuracy; however, the algorithms were only tested in binary classification, and multiclass classification was neglected. Also, out of five types of attacks in the BoT-IoT dataset, only one was used in the study.

Seven algorithms were compared in [25] by accuracy and detection rate to single out which had the lowest false alarm. Both the BoT-IoT dataset and the CSE-CIC-IDS2018 dataset were used. The algorithms compared were: RNNs, deep neural networks (DNN), restricted Boltzmann machines (RBM), deep belief networks (DBN), convolutional neural networks, deep Boltzmann machines, and deep auto encoders. While the findings of this study are acceptable, not cleaning the datasets and priming them before using them in the models gives way to inaccuracies in the results.

An important step in the detection process is the pre-analysis of data before feeding it to the classification algorithms, a process known as feature selection. In [26] and [27], different hybrid feature selection methods were utilized to find the minimum number of features with the best results, in which normal traffic and attacks can easily be distinguished with less time and effort.

*B. AI-based detection techniques for IoT networks*

While AI and ML have had a significant impact on cybersecurity and enhancing IDSs, there is still much room for improvement, especially considering IoT networks' time constraints. To gain an edge in detection time, some IDSs have begun incorporating big data to analyze the ever-growing amount of data generated by networks. IDSs tested on traditional networks, and those that have incorporated AI and big data in their detection procedure, will be reviewed in this section.

A three-layered framework for an IDS that used DL and big data together was proposed [28]. The data would first be entered for analysis, then would be used in a cascade learning algorithm, and finally, a DL algorithm called Multilayer Perceptron (MLP) would detect the attacks. Evaluation of this framework was based on accuracy and F-measure. Although the framework was claimed to have handled a massive load of data in minimal time, the results show that neither the accuracy rate nor the F-measure achieved more than 75%.

The work of [29] firmly stated that standard ML techniques aren't strong enough to detect some of the more advanced cyberattacks, and so presented a distributed approach to detect abnormal events in large networks. The study used DBN, multi-layer ensemble SVM, and Apache Spark to apply this distributed approach, and used Area under Receiver Operating Characteristic, precision, recall, f-measure, and training time to evaluate it. And while the distributed model performed extremely well in detecting anomalies in the network, the training time it had to go through beforehand was much longer than the models it was compared against in the study.

For IDSs to be applicable in an IoT setting, they must be able to detect attacks in real-time. To achieve this, paper [30] proposed a framework that combines the CC4 neural network proposed in the study of [20], ML, and finally, real-time data analysis using Apache Storm, which helped reduce training time substantially. The criteria used when testing the framework consisted of accuracy, FPR, training time, and FNR. The framework shows promise in accuracy and FPR, in which it achieved 89% and 4.32%, respectively, but unfortunately fell below 90% when it came to average accuracy.

Studies [31] and [32] have both sought to use Apache Spark to detect DDoS attacks in large traffic volumes. Only [31] combined signature-based with Artificial Neural Network (ANN) algorithm, while [32] used only the ANN algorithm. While the results from both studies show high accuracy, the detection models and tests were limited to a single attack.

Some IoT network attacks happen from within the network itself; to detect these kinds of attacks, special attention must be given to heterogeneous data within the network. Paper [33] focuses on this scenario and uses Apache Hadoop and Apache Spark to analyze and classify the traffic within the network, and then evaluates the tests done based on accuracy, FPR, and FNR. The results for all three criteria proved favorable; however, these three criteria alone aren't enough to prove the model's effectiveness.

One of the features big data environments offer is their extensive AI and ML libraries. Gupta et al. [34] leverages Apache Spark's MLlib to present an IDS framework that cuts down on time and processing power using feature selection. Two feature selection methods were tried and compared: correlation-based and chi-squared, both of which mined the features from DARPA's KDD'99 dataset, and also from its enhanced and renewed version, the NSL-KDD dataset. These features were then used by five ML algorithms: regression, SVM, RF, NB, and Gradient Boosted DT. To evaluate these algorithms and features, training time, prediction time, accuracy, sensitivity, and specificity were used, the results being that not one version of the IDS could detect with an accuracy of more than 93%, and some algorithms didn't even make 36%. Moreover, accuracy rates in general dropped when using the NSL-KDD dataset instead of the KDD'99, implying that the IDS might not be suitable for detecting more modern attacks.

Authors in [35] implemented an IDS that should be capable of processing a large amount of data in a very limited time. The system was built using Apache Hadoop and Apache Spark based on the findings of Hive SQL and unsupervised learning algorithms. It was then tested using the UNB ISCX 2012 dataset, a dataset by the Information Security Centre of Excellence, a cyber-security lab at the University of New Brunswick. When tested, the IDS could only detect attacks with an accuracy rate of 86.2% and a FPR of 13%, rates that are not acceptable in real security settings.

Another work that relies on Apache Spark and feature selection is [36], which used Apache Spark along with Hadoop MapReduce to build an IDS. For the feature selection, the Non-dominated Sorting Genetic Algorithm is used to ascertain the features from the NSL KDD'99 dataset. The results were acceptable: an accuracy of 92.03%, a detection rate of 99.38%, and a testing time of 0.32 seconds. However, the dataset used in this study is very old and does not contain modern attacks, some of which could be used to target IoT networks.

### C. Big data detection techniques for IoT networks

While the previous section reviewed research that combines AI and big data for IDS uses in traditional networks, this section narrows the scope further, reviewing researches that do the same in IoT network environments.

Despite the clear connection IoT, AI, and big data have in terms of functionality and business uses, sources are scarce when it comes to the involvement of AI and big data together in cybersecurity measures for IoT networks. Big data is the main factor in analyzing and monitoring IoT networks, without which these networks would lose their value in generating insights from the huge amount of data they collect. However, given IoT's recent appearance, the rarity of large IoT datasets, and the complications in setting up distributed systems for big data platforms, research in big data for IoT cybersecurity is still lacking. Studies relatively close to this sector either focus on joining AI and big data in general or focus on joining AI and IoT without mention of big data, marking a pressing need for researchers to take a closer look at the issue. After meticulous searching, three works were found that combined these three elements together in one way or another.

A big data framework for intrusion detection was presented in [37] and implemented using Apache Spark as a platform. To detect attacks, the framework used multiple classifiers: DNN, SVM, RF, DT, and NB. After testing the framework using a raw dataset, it was concluded that the DNN classifier had the highest accuracy; however, compared to other classifiers it also had the highest prediction time, and the accuracy rate was less than 80%.

Vinayakumar et al. [38] implemented an IDS using big data and DL algorithms MLP and FNN. The work was done using Apache Spark and Apache Hadoop Yet Another Resource Negotiator. The evaluation criteria were accuracy, precision, recall, f-measure, TPR, and FPR. Results indicate that the model was indeed more powerful than the usual IoT IDS; however, the accuracy rate for detecting some attacks using multiclass classification dropped below 90%.

To chiefly detect SYN-DoS attacks in IoT networks, the author in [39] came up with a big data framework using Apache Spark and a number of supervised learning algorithms included in Spark's MLlib. It was concluded that the framework did indeed have high accuracy in a relatively short amount of time, but as pointed out, the framework was designed with only one attack in mind.

### III. METHODOLOGY

#### A. Overviw

In this section the methodology of the proposed IDS system and tests is explained. The methodology starts with a concise and detailed view of relevant literature to gain insights, find the current concerns of the research community, and define gaps and weak points that may be filled or strengthened. The next step is setting up a suitable environment to withstand large amounts of data and the analysis of that data, then reworking that data and cleaning it to gain more accurate results. Next is the development of two models for intrusion detection—the first for detecting any attack no matter the kind (binary classification), and the second for detecting each known attack with impeccable accuracy (multiclass classification). Finally, both models will be evaluated by standard IDS performance metrics.

The methodology adhered to in this paper can be summarized as Fig 2 illustrated.

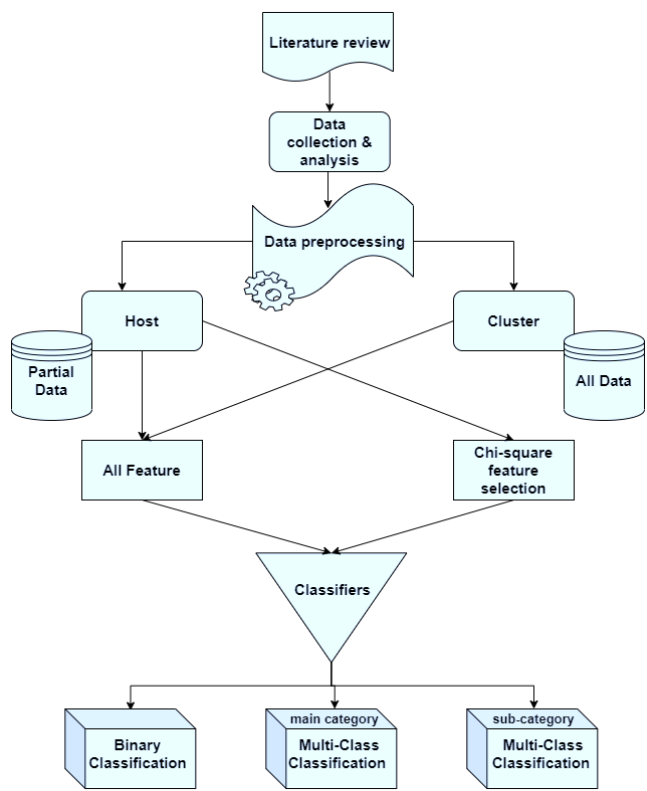

*Fig 2 Methodology*

## B. Dataset

Given that IoT networks are relatively new, the dataset needed in this paper for training and testing must contain new types of attacks along with old types of attacks, as both may be used to attack IoT networks. The data must also function as a realistic testbed simulating the IoT network environment, with data large in size and varying in type. Both of these criteria are met in the BoT-IoT dataset [40], which contains different types of traffic flows such as normal traffic, IoT data, and several IoT specialized attacks such as the botnet attack. To make good use of the dataset in detecting different attacks and classifying them, the data has been mined for the most important features to be used in AI and ML methods. The data has also been labeled as attack and non-attack, and the attack section has been labeled further into categories and subcategories to make the data usable in researching multiclass classification.

Dealing with the BoT-IoT dataset was one of the big challenges in this paper. Due to its large volume, the authors in [40] divided the data into 74 CSV files, each of which contained one million records. In order to deal with this large number of files, Apache Spark was used to combine all of the files into one CSV file. Since the dataset is too large to fully list.

Opening all these files was not possible without combining them, and Apache Spark uses techniques that minimize memory usage to avoid overloading the system. The union function provided by Apache Spark gave the ability to combine two data frames into one data frame.

What remained was only to download all the files to the machine, then upload them to an Apache session using the spark.read function. There the files are combined and saved as one file. Reading and saving the files were made easy by using Spark's read and write function, which gives many options such as file formats, inheritance schemas, and headers.

## C. Data preprocessing

When dealing with large-scale datasets, it is often noticed that some of the data is in disarray, missing, or redundant. If data in that state is used by ML algorithms it will lead to inaccurate predictions and in turn misleading results [41]. One of the most critical steps is priming the dataset and cleaning out duplicated and unuseful records. Another issue is the variance of values throughout different features and the bias it may lead to in decision making, so normalizing the data is also crucial. However, to deal with these issues the data must not contain any strings, which is why StringIndexer was used.

The string indexer encodes all string values in a particular column into index values. six columns in the dataset contain string values which are: proto, flgs, state, sport, dport, and label.

### 1) Removing Duplicated values

Duplication is one of the most important problems faced when dealing with datasets, especially large-scale ones [59]. This is due to the increased processing power and training time needed for these duplicated records, and it may also affect some algorithms' results negatively rather than just prolong them. To remove duplication in the BoT-IoT dataset, an Apache Spark built-in method for duplicate removal "dropduplicates( )" was used on the dataset and returned that zero duplicates were found.

### 2) Remove missing and unkown values

Another problem faced when dealing with datasets is missing values, as records missing any values may have a negative effect on the model's accuracy [42] [10] [43] [44] [45]. There are two ways to deal with missing values in a record. The first is to fill the value in with a zero or the median of the value in related records; the second is to delete the record completely, omitting the need for assumptions. In this paper, any records with missing values will be removed, and any feature with no corresponding values in the records will not be considered. In this paper, 2833 records out of the entire dataset were discarded for having missing values, or features with no corresponding values. The distribution of these values by class are shown in Table I.

TABLE I. DISTRIBUTION OF MISSING RECORDS IN THE DATASET GROUPED BY CLASS

| No. | Class | # of missing records |
|---|---|---|
| 1 | DDoS_TCP | 499 |
| 2 | DoS_HTTP | 26 |
| 3 | DoS_UDP | 522 |
| 4 | Theft_Data_Exfiltration | 4 |
| 5 | DDoS_HTTP | 30 |
| 6 | Theft_Keylogging | 5 |
| 7 | DDoS_UDP | 420 |
| 8 | Reconnaissance_OS_Fingerprint | 128 |
| 9 | Reconnaissance_Service_Scan | 320 |
| 10 | Normal | 471 |
| 11 | DoS_TCP | 378 |

### 3) Data sampling

Another issue to consider when dealing with datasets is the inconsistent number of each class of attacks. In the BoT-IoT dataset there is a large number of DoS attacks but only a few exfiltration attacks; this imbalance can cause a bias towards attack classes with a larger number of instances. To deal with this issue, classes with too many instances will be reduced by under sampling the class to minimize the bias between classes.

Since under sampling will be used, and the ratios of instances will be strictly observed, the experiments will be conducted on two datasets. The first dataset is a partial one with under sampling, to minimize bias in the results, and also to accurately compare against other studies, since other studies use partial datasets as well. The second dataset will be the complete BoT-IoT dataset, with the purpose of testing the feasibility of big data and Artificial Intelligence in IoT security solutions.

For initial testing, the dataset was split into two parts, one containing 70% of the data and used for training, and the second containing 30% of the data and used to test the model. After achieving favorable results in the initial testing phase, k-fold cross-validation with k equal to 10 was used. K-fold validation was selected to guarantee samples aren't over fitted with low variance and bias, and that they also have good error estimation.

### 4) Normalization

The last issue to deal with when priming datasets for feature selection and ML algorithms is getting rid of the bias between features caused by the different numerical representations of each value. This is done by unifying value representations under a single range using scaling and normalization. Out of the many types of normalization, for the purposes of this paper, min-max normalization was chosen with a range from 0 to 1.

### D. Feature selection

Feature selection was introduced to speed up the training and detection processes in IDSs powered by AI, but there is a tradeoff to consider when choosing features [46]. If too many features are given to the model, the training and detection times will be slowed down, defeating the purpose of using them in the first place; however, if the number of features is too low, the accuracy of the model will be compromised. Another consideration to make when feature selection is used, is the method of selection itself. Machine learning models have different types of inputs and outputs, namely numerical and categorical, with each having a more suitable approach for feature selection. In our case, the inputs and outputs will are categorical, and so, the chi-square feature selection method will be used to choose the crucial features for the model. The method relies on the chi-square metric used to calculate the differences in categorical variables, then uses hypopaper testing for evaluating the results. Chi-square feature selection is included in Apache Spark's MLlib, making it easy to integrate into the model.

### E. Classification based Intrusion Detection Schemes

To properly achieve this study's objectives, it must be ensured that data is correctly classified under the pressure of big data, whether it be classified as safe or harmful, or data already known as harmful and the type of attack it belongs to is identified [47]. While binary identification can protect systems from immediate threats, categorizing incoming attacks into different classes can also prepare for the future by showing the frequency of recent attacks and enabling administrators to apply the appropriate security measures. This benefit only increases the more classes we categorize into. So instead of only testing main class classification, two columns of the dataset were used to create subclasses, and subclass classification was tested as well. Of the papers previously reviewed dealing with big data, the three most common classifiers will be chosen and compared to gain a wider view and better results. It is worth noting that of the reviewed literature, no paper was found that attempted the additional step with subclasses and their classification, and so, no comparison of results could be made.

### F. Apache Spark setup

For this research to meet all its goals, the environment used should be specialized for containing and analyzing big data, while remaining fast, easy to use, and robust. These conditions are all met by Apache Spark, which has an additional feature—MLlib, which contains many methods and ML algorithms needed to build the model. Apache Spark supports multiple programming languages such as Scala, Python, Java, and R. In this paper, Python will be used due to its capabilities, ease of use, and large community, while Apache Spark will be set up on Google Cloud Platform.

### G. Model evaluation

To demonstrate the use of the model and findings gained from testing, known and agreed-upon evaluation metrics must be used, such as accuracy, precision, recall, and f-measure. F-measure was chosen out of these metrics since it is the most suitable for unbalanced data [48]. Apache Spark calculates these metrics using its evaluation library, but it only supports RDD input, so results must be converted to RDD before calculation.

## IV. PARTIAL DATA

This section will detail the experiments and results on a subset of the BoT-IoT dataset that was derived from the original dataset using an under sampling technique, which randomly selects instances from each class at fixed ratios to eliminate bias in the data. The ratios that were chosen in this paper are based on correcting the high imbalance in the original dataset by assigning a higher ratio to classes with fewer instances and a lower ratio to classes with more instances.

In total, 27 experiments have been made on this partial data using one virtual machine instance that was set up on Google Cloud with 4 vCPUs and 15 GBs of RAM. The virtual machine carried out the experiments using Ubuntu 18.04, Python 3.7.6, and Apache Spark 2.3.0. The experiments focused on the three most commonly used classifiers in related works, namely: RF, DT, and NB. No modifications were made to the default parameters documented in Apache Spark's classifier documentation and referenced in [19]. Sample codes of some experiments are presented in appendix A.

In the following sections, three main divisions will be explored, the first being binary classification, where attacks are represented as ones, and the rest of the traffic is represented in zeros. The second part is main category multiclass classification, where attacks are not just detected indiscriminately but rather classified into one of four main categories of attacks to better deal with each one; adding classifying normal data traffic as well makes it a five-class classifier. The final part is subcategory multiclass classification, where classification goes deeper and identifies 11 classes based on 10 attack subcategories and normal traffic. Going through each of these three divisions, f-measures for all the classifiers will be reviewed and compared to identify the top performing one, and then compare f-measures for all best-case scenarios at the end.

### A. Binary Classification

The first subsection in this series of tests is concerned with binary classification, where attacks are labeled positive, or 1, while normal traffic is negative, or 0. From a security point of view, binary classification is the most important of classifications, for we can surely gain much by identifying the type of attack, but the priority is to detect this attack and prevent it from compromising the system. In the following tests, 29,507 instances of attacks and 2761 instances of normal data are taken from the partial dataset.

Overall, nine tests were conducted; these are divided into three groups based on the number of features used by the classifier, where the first group uses all the features offered by the dataset, the second uses the top 10 features, and the third uses only the top five features. The purpose of reducing features is to improve performance while maintaining high accuracy, for accuracy is the focus in this paper, and performance is handled by the big data platform and therefore secondary.

*1) All Features*

Three tests were conducted on the partial dataset, each test for a specific classifier in order to binary classify incoming traffic. Of the three classifiers, RF had the best results with only one FN and 41 FPs. These results are quite astonishing because from a security perspective, FN represents that an attack has evaded intrusion detection. That's why it is considered the most important aspect of the matrix because it is better to stop legitimate traffic than to pass an attack to the network. All other data was correctly classified. A close second was the DT, which considered 29 attacks as normal traffic and had 44 FP instances. The real problem is NB, which may have had only three FNs, but also had 2674 FPs, which means a lot of normal traffic will be classified as attacks and blocked, causing a DoS attack. This is almost always the case with NB in the upcoming tests, indicating its theory isn't suitable for the settings presented in this paper.

To minimize bias due to the high imbalance of the dataset, plus to gain a comprehensive overview of accuracy, precision, and recall all in one reading, average f-measure was used, as shown in Fig 3. The average f-measure was an astounding 99.6% for RF and 99.3% for DT, while NB scored only 50.9%. This indicated that RF and DT classifiers are reliable in binary classification, with RF slightly outperforming DT in the tests.

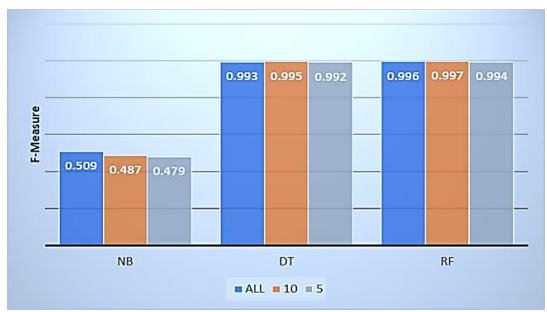

Fig 3 Binary classification All Cases

*2) Top 10 features*

As discussed in the methodology, chi-squared is used as a feature selector for all upcoming tests. This is done by selecting the top features to classify by using the topnum feature with either 5 or 10 features as the output. In this section we'll start with 10 features.

It can be found that although the features now used are less than half of the original features, not only has accuracy been maintained, but it has also improved in some cases. In RF, FNs have almost doubled from four to seven, while FPs have decreased from 41 to 28. However, for DT, both FPs and FNs have decreased—FPs from 29 to 19 and FNs from 44 to 31—showing fewer features actually resulted in better performance in DT. Lastly, NB seems even worse off with fewer features, as in the previous test normal traffic labeled as attacks were very high, while only 15 instances were correctly predicted as normal.

Fig 4 shows the average f-measure for each classifier, where RF and DT both scored 99.7% and 99.5%, respectively, which is higher than the results in the tests made with all features, indicating more features don't always lead to better performance.

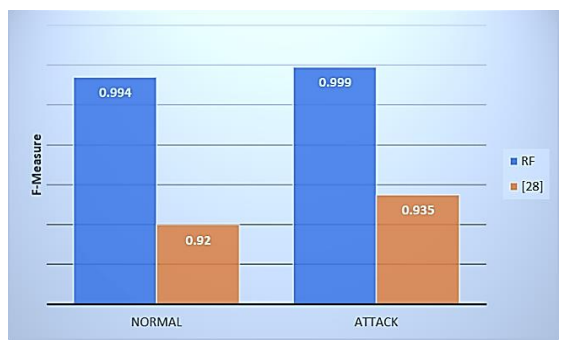

Fig 4 Best obtained results vs. results obtained by [28] (Binary).

*3) Top 5 features*

The final group of tests is conducted using only five features. This decrease in features may increase performance but will cause a tradeoff in accuracy, and while that tradeoff happened in the tests as expected, accuracy did not decrease by much. In general, RF saw a 50% increase in mistakes while DT saw less of an increase, however NB classified all packets as attacks except four, marking its worst result yet. Average f-measures slightly decreased compared with previous tests, as shown in Fig 4.

*4) Summary*

In order to determine the best classifier and number of features, all results from the previous tests are shown in Fig 4. We can see that RF scored the highest f-measure of 99.7% when using 10 features in binary classification, and all other scenarios with RF and DT did not fall below 99.2% regardless of the number of features they use. Meanwhile, NB lagged behind, its best average f-measure not exceeding 51%. To see the results for the best scenario in more depth, Fig 5 shows the detailed f-measure for the best-case scenario, which is RF with 10 features. As shown, the f-measure for attacks is 99.9%, for normal instances 99.4%, and the average is 99.7%. And while the average f1 would remain the same if f1 for attacks and normal instances were swapped, these results are considerably better from a security perspective, since it is safer to assume a normal packet as an attack than vice versa.

To gain a better understanding of the results of this research, the work of [40] was chosen to compare results against. This paper was chosen due to its use of AI and IoT to implement binary and multiclass classification scenarios using the BoT-IoT dataset, but where this research uses under sampling to minimize the bias resulting from the dataset's high imbalance, [40] maintains the original dataset ratios when sampling, and uses a hybrid IDS model with a C5 algorithm as its first stage and one-class SVM as its second stage. This has led to a lower attack f-measure of 93%, and a lower normal f-measure of 92%, as illustrated in Fig 5.

To summarize, three classifiers were tested for binary classification, each time with a different number of features.

Results have shown that RF and DT are very promising classifiers with higher than 99.2% average f-measures, indicating both can be used in IoT environments where accuracy or performance is the main concern. The best scenario was determined by accuracy as RF with 10 features, which scored 99.7% average f-measure, while NB had consistently insufficient results and got worse the more features were removed.

## B. Multiclass classification (Main Category)

After exploring binary classification for the three chosen classifiers in the last section, here we will go a step further by testing multiclass classification and the accuracy with which incoming traffic can be classified into five categories: normal traffic, DoS, DDoS, Reconnaissance, and theft attacks. Multiclass classification can give insights into what extra security measures can be added based on the attacks being intercepted; for example, filtering data can help decrease DoS attacks, while Reconnaissance or theft attacks can be countered by adding encryption and access control to the platform.

### 1) All Features

As in the previous section, three classifiers were tested using all features in the dataset. The binary labels of 0s and 1s were replaced with the main category feature, which was done by changing strings into doubles, then back to strings after the tests using StringIndexer(), and five classes were mapped to numbers from 0 to 4.

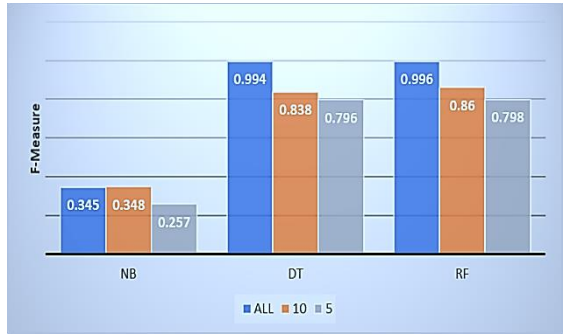

Fig 5 Multiclass classification All Cases (Main category).

The results showed unexpected high accuracy. DT and RF did exceptionally well in identifying recon, theft, and normal packets with only a couple of negligible mistakes. Only DT mistakenly labeled some DDoS attacks as DoS. In contrast, RF labeled some DoS attacks as DDoS. This is expected since both attacks have similar behaviors except for the attack source, so considering IP addresses aren't part of the features it is amazing they were classified with so few mistakes, especially since the attacks are still labeled as threats and stopped and there is no safety concern. Overall, the number of attacks considered as normal packets is 30 out of 29507.

NB wasn't as accurate as the others, as it mistakenly classified all theft attacks as different classes. Furthermore, many attacks were mistaken as normal packets, showing NB as less than adequate in multiclass classification. As for average f-measures, Fig 5 puts DT and RF over 99%, while RF had a slightly better score of 99.6% compared to DT's 99.4%. Both are acceptable f-measures in an IoT environment, however NB failed to get a passing score at only 34.5%.

### 2) Top 10 features

Following the excellent results shown by the RF and DT classifiers when using all features provided by the dataset, features were decreased to 10. It was expected that mistakes would increase, however for RF and DT, this only affected DoS and DDoS, where almost half of each were wrongly labeled as the other. At the same time, the rest of the classes were unaffected and some even saw improvements. This is not good for those who specifically want to distinguish between DoS and DDoS, whereas if both were considered of the same family, the results would be satisfactory. NB was even worse off than the test with all features, classifying normal packets as attacks leading to the possibility of a DoS attack as explained in the binary classification results. Average f-measures in Fig 6 show a 10% decline due to the misclassification of DoS and DDoS attacks with performance remaining the same, putting RF at 86% and DT at 83.8%. Notably, NB didn't exceed 50%.

### 3) Top 5 features

After decreasing the number of features to five, mistakes increased not only in DoS and DDoS, but also in reconnaissance attacks as they were labeled DoS/DDoS attacks. This was due to the fact that all three types use legitimate data as a means for exploitation, making it hard to distinguish between those with minimal features. However, normal and theft accuracies weren't affected in RF and DT. NB on the other hand deteriorated significantly, and only six normal packets were correctly identified. The rest are marked as attacks, causing almost no data to flow into the system, effectively creating a DoS attack.

As for average f-measures, there was an overall 5% putting them at 79.8% for RF, 79.6% for DT, and 25.7% for NB. These results are expected when trying to classify five classes using only five features. Fig 6 shows the average f-measure for the three classifiers.

### 4) Summary

After conducting nine tests on three groups based on the number of features, we can conclude that decreasing features has a greater impact on multiclass classification than on binary classification, as average f-measures dropped from 90% to 80% to 70% each time the number of features was halved. As shown in Fig 6, RF scored the highest average f-measure in the all features scenario, with DT using all features being a close second.

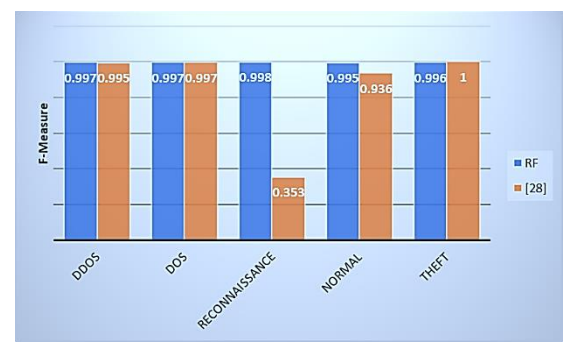

Fig 6 Best obtained results vs. results obtained by [28] (Main category)

Fig 7 lists detailed f-measures for the RF all features scenario, all of which are higher than 99.5% and with reconnaissance even being 99.8%. This marks RF as powerful in classifying across families. Since the proposed big data platform handles the load of using all available features and performance being secondary to this research, these results are satisfactory from a practical security standpoint.

The results are compared to the IDS implemented by paper [40] in Fig 7, showing that while the hybrid implementation scored a higher f-measure for theft attacks, it scored lower in every other f-measure, most notably f-measure for reconnaissance attacks. This is most likely due to the maintained ratios of the original dataset and the bias resulting from it.

Unlike binary classification, it can be concluded from these tests that the effect the number of features has on accuracy is in direct proportion to the number of classes to be classified. However, the tradeoff between performance and accuracy based on the number of features remains valid. The results in the case of all features are astounding, with a 99.6% average f1 in the best case, and no f-measure dropped less than 99.5% for each class.

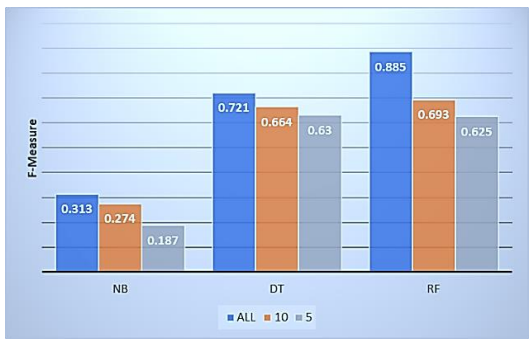

Fig. 3. Multiclass classification All Cases (Subcategory).

## C. Multiclass classification (Subategory)

After achieving excellent results in binary and main category classification, in this section we dig even deeper in an attempt to classify data into their subcategories. For example, instead of categorizing DoS and DDoS as two families, each of these is now categorized into UDP, TCP, and HTTP types.

### 1) All Features

As in previous tests, the labels for the data were replaced with subcategory labels. In the original dataset, subcategory and main category labels were separate, so for the purposes of this section, they were concatenated, resulting in 11 classes in total. After conducting the tests using all features in the dataset, good results were recorded. This is especially true for RF, which achieved high accuracy in all classes. The only exception was the theft-exfiltration attack, mostly mis-categorized as theft-keylogging due to the total instances of this specific attack being only 36, which did not provide enough training to the model compared to the 6000 instances of DoS UDP. This resulted in no classifier successfully identifying theft-exfiltration attacks, and some even misclassified them as normal data, which is expected given that the attack steals data traveling as normal packets. The rest of the classes in the test with RF saw minor to no mistakes. For example, DDoS_TCP and DDoS UDP attacks were classified perfectly, and even when mistakes were made, all attacks stayed within the same main category. DT was less accurate overall, and NB continued to show it is unreliable in the conditions this paper poses. Although RF and DT were similar in past cases, Fig 8 demonstrates that this time RF exceeded DT by 15% with an average f-measure of 88.5%, while DT scored 72%.

### 2) Top 10 features

When testing the classifiers using 10 features, DT and RF accuracy saw a large drop, making many misclassifications, and it is apparent that when the number of classes is equal to or more than the number of features, accuracy decreases substantially. RF may have managed to correctly identify normal traffic, but some attacks were mistakenly considered normal traffic, creating an easy way to exploit the system. Despite this, and the fact that the average f-measure couldn't pass 70%, RF is still considered the best case in this group of tests. Fig 8 shows the average f-measure for the three classifiers.

### 3) Top 5 features

If using 10 features was a problem for accuracy, five features only made matters worse. Signifying that 10 attacks can't be classified using five features. To add to this, many attacks were considered normal traffic, creating a significant security risk. These tests prove that the more classes we have to sort into, the more features should be used. In Fig 8, it can be seen that the best average f-measure belongs to DT at 63%.

### 4) Summary

As the results of using 10 and 5 features were not acceptable, Fig 8 shows f-measures for all the subcategory tests side by side. Notably, no average f-measure was higher than 70%, with the best case being RF with 88.5%. Considering that the classification was into 10 classes and no FNs were recorded, this can be considered a very high f-measure. This is also the first time RF and DT have a clear

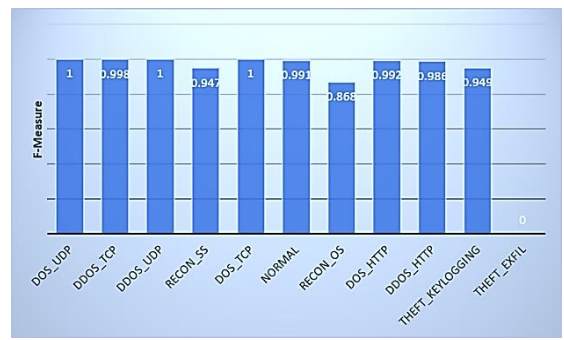

Fig. 2. Best algorithm detailed results for Subcategory classification (DT, all features).

winner, as opposed to other sections.

F-measures for each attack are displayed in Fig 9 for the RF all features scenario. The results are good overall, with all types of DoS attacks having an f1 higher than 99%, while the lowest class was recon at 86% due to its similarity to normal data, and theft-exfiltration pulling a zero only because few instances were in the dataset, leading to its mis-categorization, albeit within the same class.

After conducting all the experiments with the same criteria and methodology, it becomes more apparent that the number of classes must not exceed the number of features if high accuracy is desired, for in the previous section when classifying into five main categories, five features was the breaking point in terms of accuracy, while in this section 10 features was less than enough. As a result, the best case for multiclass subcategory classification is RF all features with an average f-measure of 88.5%.

### D. Partial Data conclusion

In conclusion, RF was considered the top-performing classifier in all sections in this section, in contrast to NB, which not only scored last in every test, but scored so low it isn't considered an option for this case of cybersecurity classification. The best case for binary classification is RF using 10 features with an average f-measure of 99.7%. RF all features is best for multiclass main category classification at an f1 of 99.6%, and finally RF all features is best for multiclass subcategory classification with an f1 of 88.5%. Trends in the data show accuracy is greatly affected if the number of features is equal to or lower than the number of classified classes. Otherwise, a tradeoff exists between accuracy and performance.

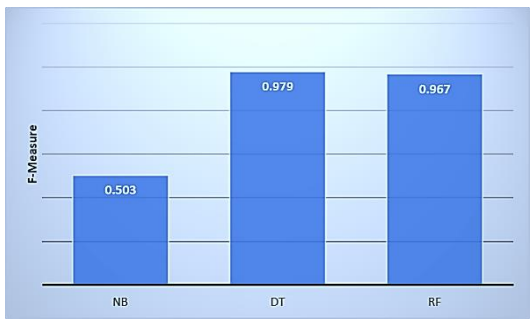

Fig. 5. Comparision of algorithms performance of the BoT-IoT dataset (Binary).

## V. FULL DATA

After testing binary and multiclass classification using a part of the BoT-IoT dataset and achieving excellent results, this section tests using the whole dataset. The BoT-IoT dataset contains over 73 million instances of attacks and normal traffic and hasn't been used in its entirety in previous research due to its significant load. Apache Spark and its big data capabilities enable us to overcome this hurdle and use the dataset's full potential. In this section the effect of large volumes of data on the model's accuracy and performance will be observed. Since the partial dataset was sampled to overcome the high imbalance in the number of attack instances, testing using the whole dataset forgoes this advantage, and biases in the upcoming tests' results are to be expected, especially given that some attacks have more than a million instances, while some have less than a thousand.

In this section, nine tests will be conducted in total, three of which will be binary classification, three others will classify into main classes, and the final three will classify into subclasses. For each of these groups, three classifications will be used: RF, DT, and NB.

Regarding feature selection, it can be found that the chi-square selector embedded in the apache spark is limited by the volume of data it processes, which makes using it in this section not applicable. since testing with all features garnered two of the best results in the previous section, and since this implementation makes use of big data solutions taking care of performance issues, feature selection was deemed unnecessary in this section.

Considering the large load this dataset can have on one instance, Amazon Emr-5.20.0 was used to distribute jobs and processes among a cluster. The cluster used in this paper is denoted as M4.xlarge by Amazon; containing eight virtual cores and 16 GB of RAM. This cluster has one master core that distributes the load to seven core instances. Hadoop distribution 2.8.5 is used for the distribution process, while Python 3 and Apache Spark 2.4.0 are used to execute the tests. This section will compare the three classifiers for each classification scenario, measure the effect big data had on accuracy and performance, and finally evaluate the repercussions the imbalance in the dataset had on the results.

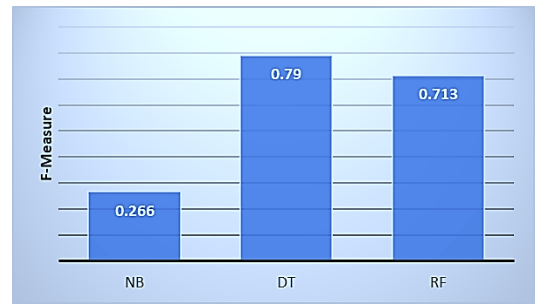

Fig. 6. Comparision of algorithms performance of the BoT-IoT dataset (Maincategory).

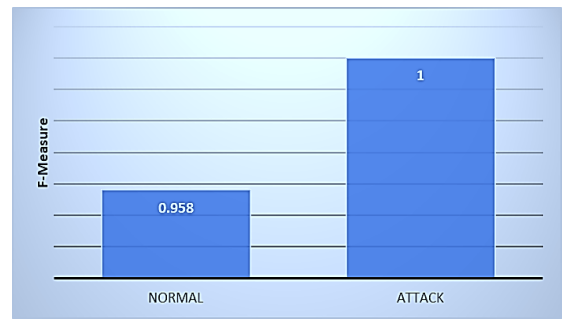

Fig. 4. Best algorithm detailed performance of the BoT-IoT dataset (Binary).

### A. Binary Classification

The first scenario that will be tested in this section is binary classification. It is considered more important than main category and subcategory classification due to its high practicality in cybersecurity settings. A challenge presented in this section is the extremely low ratio of normal instances to attack instances in the dataset, only 0.012%. Surprisingly, out of 22 million attacks, fewer than 10 were misclassified as normal traffic in both RF and DT. As for normal packets predicted as attacks, 328 and 210 are not small numbers, but taking into account how few normal instances there are in the dataset to train the model, and also that FPs are much safer compared to FNs, these numbers are acceptable. Moreover, the problem with lower accuracy in identifying normal packets can be addressed easily with more instances for training purposes. As for NB, the results show only seven correctly classified instances, marking it as unusable in the settings of this paper no matter the volume of data used.

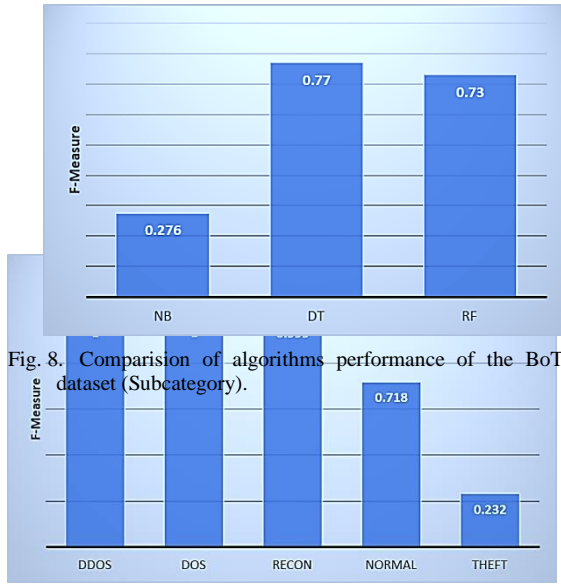

Fig. 8. Comparision of algorithms performance of the BoT-IoT dataset (Subcategory).

Fig. 7. Best algorithm detailed performance of the BoT-IoT dataset (Maincategory).

A comparison of the classifiers' average f-measure is given in Fig 10, where DT and RF are neck and neck, as in section 4. However in this section DT has a significantly higher average f-measure of 97.9% as opposed to RF's 96.7% due to scoring fewer FPs, whereas NB labeled almost all traffic as attacks causing a DoS attack, and since all attacks were correctly labeled as attacks and all normal traffic was mislabeled, NB has an average f-measure of 50%.

To have a clearer picture of the results for the best classifier, detailed f-measures will be presented as in section 5. In this case, DT is shown in Fig 11, where the f1 for attacks is 100% due to the number of mistakes being negligible, and as results are rounded to the third digit. 95% is an acceptable f1 rate for normal classification considering the aforementioned problem of few normal instances in the dataset and that losing some data is much safer than allowing some hackers into the system.

### B. Multiclass classification (Main Category)

After seeing great results in binary classification and only minor effects from the imbalance issue in the dataset, we move on to multiclass classification in this section, specifically main category classification. Main category and subcategory classifications have an added benefit over binary classification by providing system administrators with reports on the kinds of attacks repelled by the system, enabling them to react accordingly and modify security measures if needed. It is to be expected in this section that biases resulting from dataset imbalance are only more apparent due to the increased number of classes with fewer instances than necessary.

Results shows that for both DT and RF, three out of five classes are classified nearly perfectly. For example, all 11 million instances of DDoS were identified by DT. However, normal instances and theft attacks weren't classified accurately, the reason being too few instances to train the model for normal traffic, as in the previous section, and fewer theft attacks, not even exceeding 500 instances. This resulted in half of normal traffic being considered FP, causing disruption in the flow of system traffic. Theft attacks were not correctly classified either, but only two instances were false-negative while the rest were labeled as DDoS attacks. Adding this to the imbalance of the dataset with three classes having over half a million instances and two having less than two thousand, these results are understandable. NB showed the most bias in the results, wrongly mislabeling almost all normal and theft instances as other attacks and completely cutting off the flow of traffic.

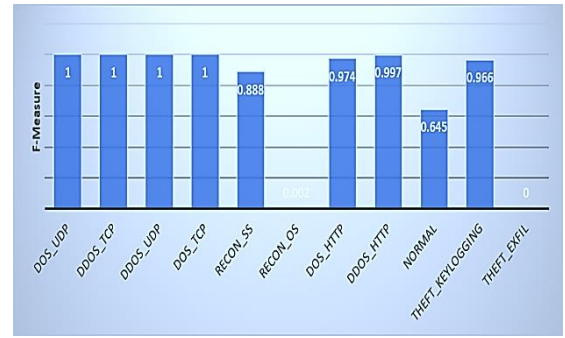

Fig. 9. Best algorithm detailed performance of the BoT-IoT dataset (Subcategory).

Accuracy is shown in Fig 12 to have decreased dramatically, but this is only because normal and theft accuracy rates were low, suggesting that increasing training instances for these two classes might substantially increase overall accuracy. The gap has widened between DT and RF with an 8% difference, where DT has a 79% average f-measure, and RF has 71.3%.

Detailed f-measures for DT are illustrated in Fig 13. The classifier has no problem with DDoS and DoS each scoring an f-measure of 100%, as well as reconnaissance attacks scoring 99%. Theft and normal f-measures were much lower, however the low f-measure for theft attacks is less of a threat when considering most theft instances were still considered attacks, in contrast to the normal f-measure signifying the loss of 29% of traffic. This makes increasing normal instances a priority in enhancing the BoT-IoT dataset or when using it to conduct ML classification research.

### C. Multiclass classification (Subategory)

The final part of this section looks to categorize traffic into a total of 11 subclasses. Detailed classifications like this, if done accurately, can help in cyber forensics and may even help in identifying the attacker. While it is hard to go through the result of each subclass, it can be noted that a large number of subclasses have high accuracy rates with the exception of normal traffic, theft-exfiltration attacks, and reconnaissance attacks. Theft attacks were a problem in the recent section, but when dealing with its subclasses, it is apparent that only theft-exfiltration attacks are problematic, and even when they are mis-categorized, they are considered theft-keylogging attacks.

A large portion of normal packets was labeled as reconnaissance attacks due to their similar nature and the imbalance of the dataset, but other than that, results in subcategory classification seem better than main category classification. Most reconnaissance attacks were successfully labeled as their main category, but most recons attacks were considered recon_ss attacks. Fig 14 shows average f-measures as low as the previous section, but this is only caused by low detection rates for normal traffic, theft-exfiltration attacks, and reconnaissance attacks.

As shown in Fig 15, DT has a 100% rate for four classes, making DoS and DDoS detection impeccable, and since reconnaissance attacks were kept in the same family even when mis-categorized, they can still be detected as the main category. Finally, the most problematic results are normal and theft-exfiltration f-measures, but when taking the imbalance problem out of the picture, these results are better than main category classification.

In this section a successful model for detecting and classifying IoT traffic using big data and ML solutions has been developed. The model was tested in binary classification, main category classification, and subcategory classification with satisfying results. DT has been clearly identified as the best classifier to be used in the case of big data. The imbalance of the dataset had a significant effect on the results, especially in main category classification. It also had an effect on the classification of normal traffic when increasing the number of classes. For this reason, it is clear normal attacks in this dataset should be reviewed and fixed before any upcoming research to avoid biases and increased FPs in the results. Finally, since this research is focused on intrusion detection and identifying attacks more than anything else, the results for this purpose are considered very good.

## VI. Conclusion

In this paper an IDS for IoT networks was developed using big data and ML. It starts with an overview of IoT networks, IDSs, and ML principles, followed by a comprehensive look into the latest research in the field, revealing a gap in IDS developed for IoT regarding the large volume of data these networks have and the effect this has on performance and accuracy. This called for the use of a big data platform such Apache Spark and ML algorithms to handle the load of IoT networks using standard criteria and methodologies. Next, the BoT-IoT dataset was selected to efficiently train the IDS due to it being relatively new, well known, and collected from a real IoT environment.

Two groups of tests were performed on the IDS. The first used a partial dataset sampled from the BoT-IoT dataset to counter the high imbalance in the number of classes it contains, and the second used the whole dataset. Each of these groups tested binary classification, main category classification, and subcategory classification using three classifiers: RF, DT, and NB. Additionally, two more scenarios were tested in the partial dataset, one using only the top 10 features, and the other using only the top five.

For the partial dataset, results show RF as the best classifier for all cases, with a 99.7% average f-measure for binary classification using 10 features, 99.6% for main category classification using all features, and 88.5% for subcategory classification using all features. For the complete dataset, DT had the best results in all cases: 97.9% for binary classification, 79% for main category classification, and 77% for subcategory classification. NB did not perform well in any of the tests.

After testing the IDS and identifying the best performing classifiers, two insights were made when looking at the results as a whole. The first insight is how greatly accuracy is affected when the number of features used in training the model was less than the number of classes to be classified. The second is the major drawback in the BoT-IoT dataset with its high imbalance and the biases it causes, especially with its lack of enough normal instances to properly train the model.